\documentclass[multphys,vecphys]{svmult}

\usepackage{makeidx}         
\usepackage{graphicx}        
\usepackage{multicol}        
\usepackage[bottom]{footmisc}
\usepackage{natbib}
\usepackage{url}
\bibpunct{(}{)}{;}{a}{}{,} 

\makeindex             

\begin{document}

\title*{The Garching-Bonn Deep Survey (GaBoDS) Wide-Field-Imaging
  Reduction Pipeline} \titlerunning{The GaBoDS Pipeline}
\author{Hendrik Hildebrandt\inst{1} \and Thomas Erben\inst{1} \and
  Mischa Schirmer\inst{2} \and J\"org~P.~Dietrich \inst{3} \and Peter
  Schneider \inst{1}} \institute{Argelander-Institut f\"ur Astronomie,
  Auf dem H\"ugel 71, 53121 Bonn, Germany
  \texttt{hendrik@astro.uni-bonn.de} \and Isaac Newton Group of
  Telescopes, Apartado de correos 321, 38700 Santa Cruz de La Palma,
  Tenerife, Spain \and European Southern Observatory,
  Karl-Schwarzschild-Str. 2, 85748 Garching, Germany }
\maketitle

\begin{abstract}
  We introduce our publicly available Wide-Field-Imaging reduction pipeline
  THELI. The procedures applied for the efficient pre-reduction and
  astrometric calibration are presented. A special emphasis is put on the
  methods applied to the photometric calibration. As a test case the reduction
  of optical data from the ESO Deep Public Survey including the WFI-GOODS data
  is described.  The end-products of this project are now available via the
  ESO archive Advanced Data Products section.
\end{abstract}

\section{Introduction}
\label{hil:sec:1}
Over the past years our group has developed a wide-field-imaging
reduction pipeline called THELI which is now publicly
available.\footnote{ftp://ftp.ing.iac.es/mischa/THELI/} Since data
reduction for most optical and near-infrared cameras is very similar
the pipeline was designed in an instrument-independent way from the
beginning. It is easily adaptable to new cameras by use of instrument
configuration files. Nearly fully automatic processing with little
need for interaction quickly leads to scientifically exploitable
results making the reduction of one night of wide-field-imaging data
(e.g.  from WFI@ESO/MPG2.2m) possible in a few hours.  Large parts of
the pipeline are parallelised increasing speed in presence of a
multi-CPU machine. A graphical user interface (GUI) is available which
facilitates the handling of reduction and configuration of THELI. A
detailed description of the pipeline can be found in
\cite{hildebrandt:2005AN....326..432E}. The modules for absolute
photometric calibration which are not part of the public release at
the moment are described in \cite{hildebrandt:2006A&A..Hildebrandt}.

\section{Structure}
The pipeline is based on a number of excellent existing open source
software packages like the \emph{LDAC} tools, different \emph{TERAPIX}
packages (\emph{SExtractor}, \emph{SWarp}, etc.), \emph{Eclipse,
  Astrometrix}, and \emph{IMCAT}, besides others. Bash scripts are
wrapped around these packages in order to handle the communication
between the different tasks, to control configuration parameters, and
to produce some plots for quality control. Due to this modular
structure the pipeline is easily extensible and modules can be
exchanged if better ones become available. This was done several times
in the past; e.g., for resampling and coaddition \emph{drizzle} was
replaced by \emph{SWarp}.

Building up a pipeline from many different software packages naturally
has some disadvantages when compared to a homogeneous system which is
developed from scratch. The data flow is not as transparent and error
handling becomes more complicated. For very large projects like the
major upcoming imaging surveys it would be desirable to be able to
track the history of each of the many thousand reduced images back to
the raw images. This can only be done with a sophisticated database
system which is at the moment not implemented in THELI.

\section{Processing}
\subsection{Pre-Processing}
\begin{figure}
\centering
\includegraphics[angle=-90,width=\textwidth]{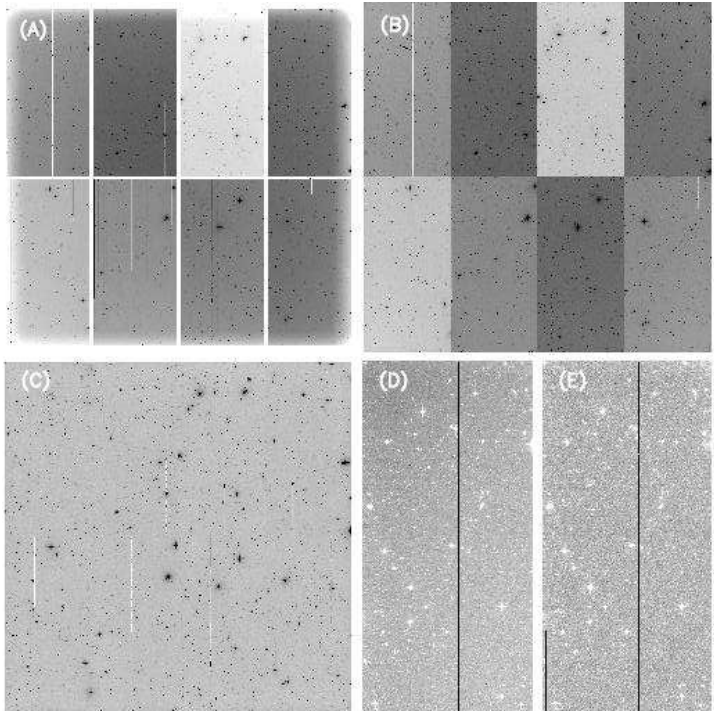}
\caption{Pre-reduction steps for WFI $V$-band data: (A) raw data,
  (B)\&(D) debiased and flatfielded data (C)\&(E) superflatted data.
  Taken from \cite{hildebrandt:2005AN....326..432E}.}
\label{hil:fig:1}
\end{figure}

The pre-processing, i.e. the removal of instrumental signatures from
the data, is done on a single chip basis and does not differ from well
established procedures applied for single-chip cameras. A standard
debiasing and flatfielding is combined with a superflat and, if
necessary, a fringe-removal.  See Fig.~\ref{hil:fig:1} for a visual
impression of some WFI data at different reduction steps.

\subsection{Astrometric Calibration}
THELI was designed with weak-lensing applications in mind. Therefore, a highly
accurate astrometric calibration was mandatory from the beginning to minimise
the impact of coaddition on the shape of the PSF. \emph{Astrometrix} is used
in combination with overlapping astrometry between all chips entering a
coaddition to achieve an internal astrometric accuracy of a tenth of a pixel.
The external accuracy is obviously limited by the accuracy of today's
astrometric standard star catalogues.

\subsection{Photometric Calibration}
The different chips of a camera are brought to the same photometric zeropoint
by division with the appropriately rescaled superflat. Since some residuals
may be left after this procedure we perform an internal photometric
calibration from overlap objects between different chips of different
exposures to derive relative zeropoints, ${\rm ZP}_{{\rm rel,}i}$, which
satisfy the additional condition $\sum_i {\rm ZP}_{{\rm rel},i}=0$.

\begin{figure}
\centering
\includegraphics[width=\textwidth]{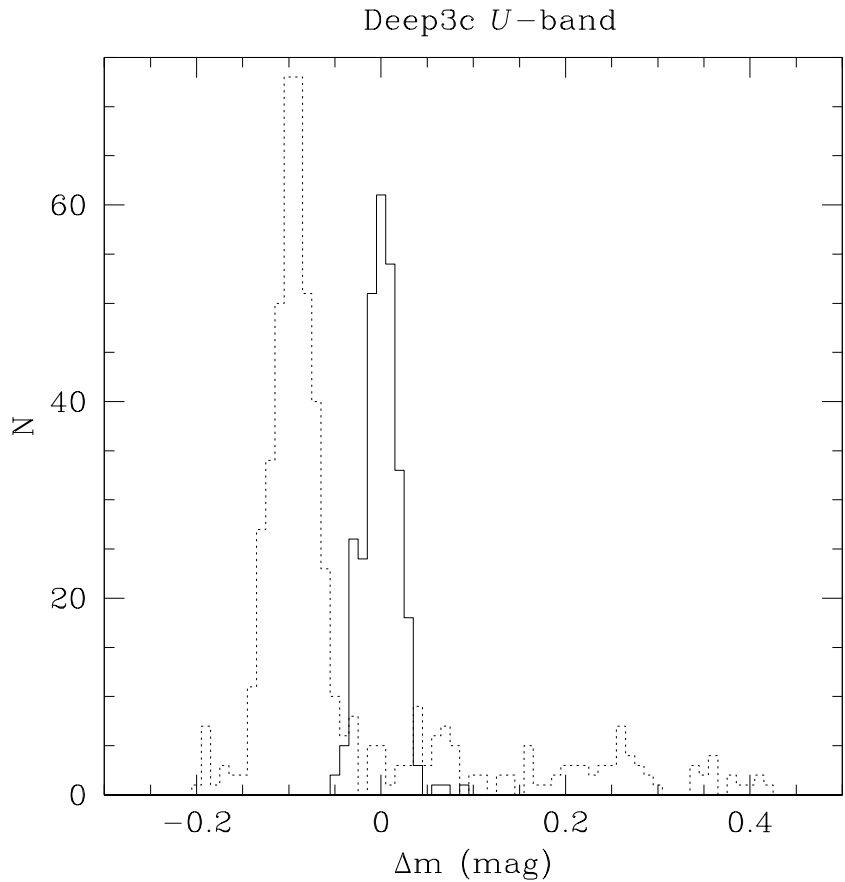}
\caption{Distribution of the corrected zeropoints (see text) before
  (\emph{dashed line}) and after (\emph{solid line}) rejection of
  apparently non-photometric nights for the calibration. Taken from
  \cite{hildebrandt:2006A&A..Hildebrandt}.}
\label{hil:fig:2}
\end{figure}

All available Landolt/Stetson standard-star exposures of the considered nights
are reduced in the same way as the science exposures. Absolute photometric
zeropoints, colour terms, and extinction coefficients are estimated from these
exposures. From those photometric parameters and the relative zeropoints we
calculate corrected zeropoints, ${\rm ZP}_{{\rm corr},i}$, for all images
belonging to calibrated nights:
\[
{\rm ZP}_{{\rm corr},i}={\rm ZP + Airmass\cdot EXT + }{\rm ZP}_{{\rm rel},i}
\]
Theoretically these corrected zeropoints should all coincide. A plot showing
the distribution of all corrected zeropoints is therefore an excellent tool to
identify non-photometric nights that should not be used for absolute
calibration (see Fig.~\ref{hil:fig:2}).

\begin{figure}
\centering
\includegraphics[width=\textwidth]{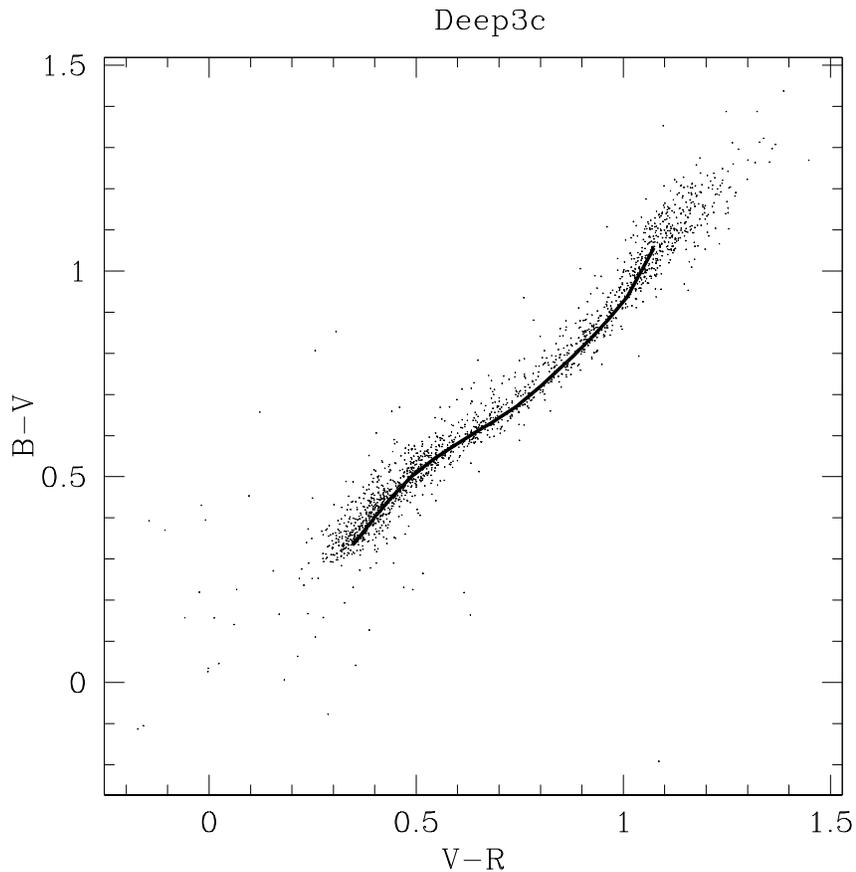}
\caption{Colour-colour diagram for stars in the field Deep3c in
  comparison to theoretical isochrones from
  \cite{hildebrandt:2002A&A...391..195G}. Taken from
  \cite{hildebrandt:2006A&A..Hildebrandt}.}
\label{hil:fig:3}
\end{figure}

Moreover, we check the absolute photometric calibration by means of galaxy
number counts and, if images in more than one filter are available, by means
of stellar colour-colour diagrams (see Fig.~\ref{hil:fig:3}).

\section{Reduction of the ESO Deep Public Survey}
The ESO Deep Public Survey (DPS) is a deep multi-colour imaging survey
overlapping with GOODS-South and carried out with WFI@ESO/MPG2.2m and
SOFI@NTT. Here, we concentrate on the optical part in the filters
$UBVRI$. The survey covers three square degrees in total of which
approximately two square degrees have full five-colour coverage to
considerable depth. The reduction, quality control, and a comparison
to a different reduction by the ESO Imaging Survey team is covered in
\cite{hildebrandt:2006A&A..Hildebrandt}.

\subsection{Survey Overview}
In Table~\ref{hil:tab:1} the main characteristics of the DPS fields are
summarised. The 5$\sigma$ limits in circular apertures of $2''$ diameter are:
${\rm mag}_{{\rm lim},U}= 25.3$, ${\rm mag}_{{\rm lim},B}= 26.2$, ${\rm
  mag}_{{\rm lim},V}= 25.8$, ${\rm mag}_{{\rm lim},R}= 25.3$, ${\rm mag}_{{\rm
    lim},I}= 24.3$)
\begin{table}
  \caption{Positions and available colours of the twelve DPS fields and the two mispointings (Deep1e and Deep1f). Taken from \cite{hildebrandt:2006A&A..Hildebrandt}.}
\label{hil:tab:1}
\begin{tabular}{ r c c c c }
\hline
\hline
field & RA [h m s] & DEC [d m s]& avail.  & comments \\
      & J2000.0    & J2000.0    & colours &\\
\hline
Deep1a     & 22:55:00.0 & $-$40:13:00 & $UBVRI$ & \\
     b     & 22:52:07.1 & $-$40:13:00 & $UBVRI$ & \\
     c     & 22:49:14.3 & $-$40:13:00 & $VRI$   & \\
     d     & 22:46:21.4 & $-$40:13:00 & ---     & \\
Deep2a     & 03:37:27.5 & $-$27:48:46 & $R$     & \\
     b     & 03:34:58.2 & $-$27:48:46 & $UBVRI$ & \\
     c     & 03:32:29.0 & $-$27:48:46 & $UBVRI$ & centred on GOODS-S\\
     d     & 03:29:59.8 & $-$27:48:46 & $R$     & \\
Deep3a     & 11:24:50.0 & $-$21:42:00 & $UBVRI$ & \\
     b     & 11:22:27.9 & $-$21:42:00 & $UBVRI$ & \\
     c     & 11:20:05.9 & $-$21:42:00 & $UBVRI$ & \\
     d     & 11:17:43.8 & $-$21:42:00 & $BVRI$  & \\
\hline                                          
Deep1e     & 22:47:47.9 & $-$39:31:06 & $URI$   & \\
     f     & 22:44:58.4 & $-$39:31:54 & $I$     & \\
\hline
\end{tabular}
\end{table}

The reduction of the DPS was the first application of our pipeline to
a large dataset consisting of more than 3000 raw science frames.
During this reduction the pipeline was extensively debugged and
optimised so that it is now in a very stable state.

\subsection{Data Release}
The data were released to the scientific community after reduction and careful
checking. The images can be retrieved from our
server\footnote{\url{http://marvin.astro.uni-bonn.de/DPS/}} or via the ESO
archive Advanced Data Products
section\footnote{\url{http://archive.eso.org/archive/adp/GaBoDS/DPS_stacked_images_v1.0/}}.
The DPS optical data are used by our group for studies of Lyman-break galaxies
at $z\sim3$
\citep{hildebrandt:2005A&A....Hildebrandt,hildebrandt:2007A&A...462..865H} and
for weak-lensing studies supported by photometric redshifts
\citep{hildebrandt:2006astro.ph..6571H,hildebrandt:2007A&Asubmitted}.

\bibliographystyle{aa}
\bibliography{hildebrandt}

\printindex
\end{document}